\documentclass[epj,referee]{svjour}
\usepackage{graphicx}
\usepackage{subfigure}

\newcommand{\bb}{\bibitem}

\begin{document}
\title{Mott Insulator to Superfluid Phase Transition in Bravais Lattices via the Jaynes-Cummings-Hubbard Model}
\author{C. B. Gomes \and A. M. C. Souza \and F. A. G. Almeida}                     
\institute{Departamento de Fisica, Universidade Federal de Sergipe, 49100-000 Sao Cristovao-SE,
Brazil}
\date{Received: date / Revised version: date}
%
\abstract{The Properties of the Mott insulator to superfluid phase transition are obtained through the fermionic approximation in the
Jaynes-Cummings-Hubbard model on linear, square, SC, FCC, and BCC Bravais lattices,
For varying excitation number and atom-cavity frequency detuning.
We find that the Mott lobes and the critical hopping are not scalable only for the 
FCC lattice. At the large excitation number regime, the critical hopping is scalable for all the lattices and it does not
depend on the detuning.} 
\maketitle
\section{Introduction}
Cold-atom systems have been regarded as efficient simulators of quantum many-body physics \cite{BDZ,JSGM}
due to its ease of controllability. Research involving ultracold bosonic systems 
has brought lot of interest in the subject \cite{AMC1,MOTT}.
Experiments with optical lattices in three dimensions have revealed 
the superfluid-Mott insulator (MI-SF) phase transition \cite{MG}.
Such phenomena have been studied in the frameworkof the 
Jaynes-Cummings-Hubbard Model (JCHM) \cite{Na1,Na2,NPA,HASP}.

The JCHM has been a widely used tool for the investigation of many-body
systems describing the interplay between the atom-cavity coupling and the 
inter-cavity hopping of photons \cite{SB}. In the absence of the hopping term, the model
reduces to the Jaynes-Cummings Model \cite{JC,EPJ1} which can be 
exactly solved within the Rotating Wave Approximation. 
When the hopping term is non-zero the solution becomes non trivial. 
The difficulty to find analytical solutions for the model has forced researchers
to resort to approximation or numerical methods for dealing with such problems \cite{FWG}.
Recently, Mering et al. \cite{AMPK} have proposed an approach in which spin
operators are mapped to the fermionic ones, hence allowing the application of a Fourier 
transform that decouples the Hamiltonian into independent
ones, which are associated to each momentum value. 
The great advantage of this method is the simplicity
in which physical quantities are found, such as the energies and the chemical potential. 
The approach presented by Mering et al. includes all classes of Bravais structures. 
However, they only derived the results for one-dimensional lattices.

Despite the experiments involving optical structures in three dimensions, 
only a few theoretical results regarding these lattices in dimensions greater than one is available 
in the current literature \cite{EPJ2}.
It lacks a systematic presentation of the phase diagram of typical Bravais lattices.
This is the purpose of the present paper. Here, we investigate the JCHM for different Bravais lattices in one, two and three
dimensions and analyse the influence of the topology on the MI-SF phase transition. We use the approach
introduced by Mering et al. \cite{AMPK}. 

The paper is organized as follows. In Section II we introduce the JCHM.
In Section III we present the fermionic approximation. Results are presented
in Section IV. Finally, IN Section V, we summarize our
main results and conclusions.

\section{Jaynes-Cummings-Hubbard Model}
The JCHM hamiltonian for a lattice of $L$ atoms is given by
 ($\hbar=1$)
\begin{eqnarray}
\hat{H} & = \omega &\sum_{j}\hat{a}_{j}^{\dagger}\hat{a}_{j}+ \epsilon \sum_{j}
\hat{\sigma}_{j}^{
+}\hat{\sigma}_{j}^{-}+g\sum_{j}(\hat{a}_{j}^{\dagger}\hat{\sigma}_{j}^{-}+\hat{a}_{j}\hat{\sigma}_{
j}^{+}) \nonumber \\
{} & {}
& - t \sum_{\langle ij \rangle}(\hat{a}_{i}^{\dagger}\hat{a}_{j}+\hat{a}_{j}^{\dagger}\hat{a}_{i}),
\label{E1} 
\end{eqnarray}
where $\hat{\sigma}^{\pm}=\hat{\sigma}_{x}\pm i\hat{\sigma}_{y}$ and $\hat{\sigma}_{x,y,z}$ are the
usual Pauli matrices, $\hat{a}_{j}$ ($\hat{a}^{\dagger}_{j}$) is the annihilation (creation) operator
of the light mode at the $j$th atom, $\omega$ is the light mode frequency, and 
$\epsilon$ the atomic transition frequency. The light-atom coupling
is represented by $g$, $t$ is the hopping integral, and $\langle ij \rangle$ denotes pairs
of nearest-neighbour atoms on the lattice.

When $t=0$, the Hamiltonian (\ref{E1}) is decoupled into $L$
independent Jaynes-Cummings model Hamiltonians. In this case the system has well-known eigenstates \cite{CNA}.
For $t \neq 0$, the atoms becomes coupled thus increasing the complexity of the solution, Since 
we can not write the eigenstates of the whole system as a direct product of single-cavity eigenstates.
As discussed in the introduction, an appropriate approach is the fermionic
approximation recently introduced by Mering et al. \cite{AMPK}.

\section {The Fermionic Approximation} \label{sec:fapp}
The fermionic approximation consists in replacing the spin operators by fermionic ones, i.e.,
$\hat{\sigma}^+$ ($\hat{\sigma}^-$) IS replaced by $\hat{c}^\dagger$
($\hat{c}$). In this framework, we can rewrite Hamiltonian (\ref{E1}) as
\begin{eqnarray}
\hat{H} & =
&\omega\sum_{j}\hat{a}_{j}^{\dagger}\hat{a}_{j}+\epsilon\sum_{j}\hat{c}_{j}^{\dagger}\hat{c}_{j}
+g\sum_{j}(\hat{a}_{j}^{\dagger}\hat{c}_{j}+\hat{a}_{j}\hat{c}_{
j}^{\dagger}) \nonumber \\
{} & {}
& - t \sum_{\langle ij \rangle}(\hat{a}_{i}^{\dagger}\hat{a}_{j}+\hat{a}_{j}^{\dagger}\hat{a}_{i}) .
 \label{E1a} 
\end{eqnarray}
This approximation allows to solve the model exactly, by means of a Fourier transformation.
For $t=0$, spin and fermionic operators are equivalent and then the approximation becomes exact.
Therefore, for small values of $t$ this approach turns out to be very accurate when dealing
with the JCHM \cite{AMPK}.  
 
Now we apply a Fourier transform to the fermionic and bosonic operators as 
\begin{equation}
\hat{a}_{j}=\frac{1}{\sqrt{L}}\sum_{\vec{k}}e^{-2\pi i \frac{\vec{k}\vec{R}_j}{L}}\hat{a}_{\vec{k}}, \\
\hat{c}_{j}=\frac{1}{\sqrt{L}}\sum_{\vec{k}}e^{-2\pi i \frac{\vec{k}\vec{R}_j}{L}}\hat{c}_{\vec{k}}, \label{E2} 
\end{equation}
then the Hamiltonian can be written as
\begin{equation}
\hat{H}=\sum_{\vec{k}} \left[ \omega_{\vec{k}}\hat{a}_{\vec{k}}^{\dagger}\hat{a}_{\vec{k}}+g (\hat{a}_{\vec{k}}^{\dagger}
\hat{c}_{\vec{k}}+\hat{a}_{\vec{k}}\hat{c}_{\vec{k}}^{\dagger}) +\epsilon \hat{c}_{\vec{k}}^{\dagger}\hat{c}_{\vec{k}} \right],
\label{E3} 
\end{equation}
where $\omega_{\vec{k}}= \omega$-$\nu_{\vec{k}}$ and $\nu_{\vec{k}}$ is the dispersion relation of the Bravais lattice. 

The Hamiltonian (\ref{E3}) corresponds to a sum of $L$ independent Hamiltonians $\hat{H}_{\vec{k}}$
($\hat{H}=\sum_{\vec{k}} \hat{H}_{\vec{k}}$), where each of them
is associated with a particular momentum $\vec{k}$. The ground-state energy
of $\hat{H}_{\vec{k}}$ is given by
\begin{eqnarray}
E^{n_{\vec{k}}}_{\vec{k}} & = & (1-\delta_{n_{\vec{k}}0}) \left[ n_{\vec{k}}\omega_{\vec{k}}+
\frac{\Delta+\nu_{\vec{k}}}{2}+ \right. \nonumber \\ 
{\ } & {\ } & \left. -\frac{1}{2}\sqrt{(\Delta+\nu_{\vec{k}})^2+4n_{\vec{k}}g^2} \right],
\end{eqnarray}
where the superscript denotes the excitation number and 
$\Delta \equiv \epsilon - \omega$ is the detuning between atomic trantition and light frequencies.
Notice that $\hat{n}_{\vec{k}}$ commutes with $\hat{H}_{\vec{k}}$.
For a total excitation number $N$ ($N= \sum_{\vec{k}} n_{\vec{k}}$) we have a particular configuration 
$\{ n_{\vec{k_1}},n_{\vec{k_2}},n_{\vec{k_3}},...\}$ that minimizes the total ground-state energy, $\sum_{\vec{k}}
E^{n_{\vec{k}}}_{\vec{k}}$. For $t \ll 1$, it is easy to see that
this configuration is $\{ n,n,n,...\}$ corresponding to the Mott insulator state. By increasing $t$, a quantum phase transition takes phace
and the system is driven to a superfluid state. Since $n$ is constant, the phase boundaries of the Mott lobes 
are $n$ dependent. Thus, the $n$th Mott lobe is obtained through an analysis of the particles chemical potential,
$\mu^{+}=E^{n+1}_{\vec{k}'}-E^{n}_{\vec{k}'}$, and the hole one, $\mu^{-}=E^{n}_{\vec{k}}-E^{n-1}_{\vec{k}}$ \cite{AMPK,GAS}, where
$\vec{k}'$ and $\vec{k}$ are, respectively, the values ​​that minimize and maximize these potentials. For $\mu^{+}=\mu^{-}$, the Mott lobe is
closed at the critical hopping, $t_c$, hence describing the MI-SF transition.

\section{Results}
In order to analyse the influence of Bravais lattices topology on the MI-SF transition, we study the one-dimensional (1D), square (SQ),
simple cubic (SC), body-centered cubic (BCC) and face-centered cubic (FCC) lattices. The dispersion relations $\nu_{\vec{k}}$ are,
respectively, given by \cite{kitel}
\begin{equation}
\nu_{k}^{(1D)}=-2t\cos(ka)
\end{equation}
\begin{equation}
\nu_{k_x,k_y}^{(SQ)} =-2t [\cos(k_x a)+\cos(k_y a) ]
\end{equation}
\begin{equation}
\nu_{k_x,k_y,k_z}^{(SC)}=-2t [\cos(k_x a)+\cos(k_y a)+\cos(k_z a)]
\end{equation}
\begin{equation}
\nu_{k_x,k_y,k_z}^{(BCC)}=-8t [\cos(k_x a)\cos(k_y a)\cos(k_z a)]
\end{equation}
and
\begin{eqnarray}
\nu_{k_x,k_y,k_z}^{(FCC)} & = & -4t [\cos(k_x a)\cos(k_y a)+\cos(k_x a)\cos(k_z a)+ \nonumber \\
{\ } & {\ } & \cos(k_y a)\cos(k_z a)], 
\end{eqnarray}
where $a$ is the lattice constant. For each structure we found the momentum vectors that maximizes
the hole chemical potentials and minimizes the particle ones in order to obtain $\mu^{-}$, $\mu^{+}$, 
and consequently the Mott phase boundary.

\begin{figure}
\includegraphics[width=.45\textwidth]{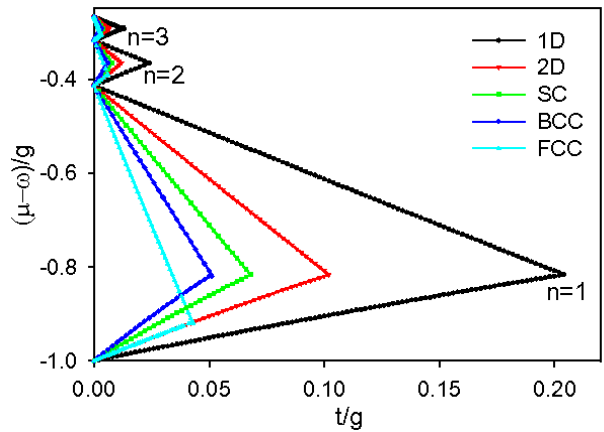}
\caption{First three Mott lobes for different lattices. Inside the lobes we have a Mott insulator state, 
while outside the system is in a superfluid state.}
\label{fig1}
\end{figure}

Figure \ref{fig1} shows the first three Mott lobes for $\omega /g=1$ and $\Delta=0$, for the considered lattices. 
We see that as the number of lattice neighbors increases, the MI phase region decreases.
This result is expected since the probability of ohoton tunneling increases for higher 
number of nearest-neighbors.
We observe that for $d$-dimensional hypercubic lattices the lobes can be rescaled by
$t_d =t/d$ causing the collapse into a single curve. 
The shape of the Mott lobes are equal for bipartite lattices, where a bipartite structure is such one 
that we can decompose into two substructures, with all nearest-neighbour sites shared between each other.
The FCC lattice is non-bipartite, Hence displaying a different behaviour. Thus
we propose to make a detailed comparative analysis between the FCC and SC lattices representing, respectively, the non-bipartite
and bipartite classes. 

\begin{figure}
  \subfigure[]{
     \includegraphics[width=0.45\textwidth]{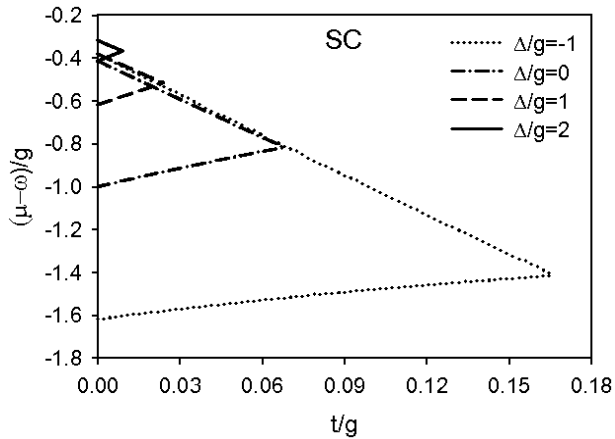}\label{fig2a}}
  \subfigure[]{
     \includegraphics[width=0.45\textwidth]{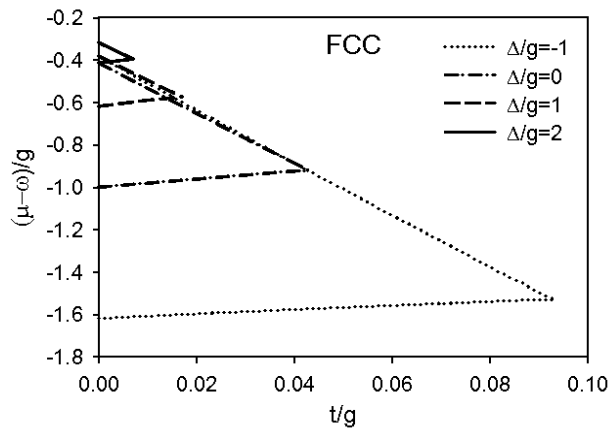}\label{fig2b}}
  \caption{Mott lobe for $n=1$ and typical values of detuning for (a) SC and (b) FCC lattices.}
\label{fig2}
\end{figure}

\begin{figure}
  \subfigure[]{
     \includegraphics[width=0.45\textwidth]{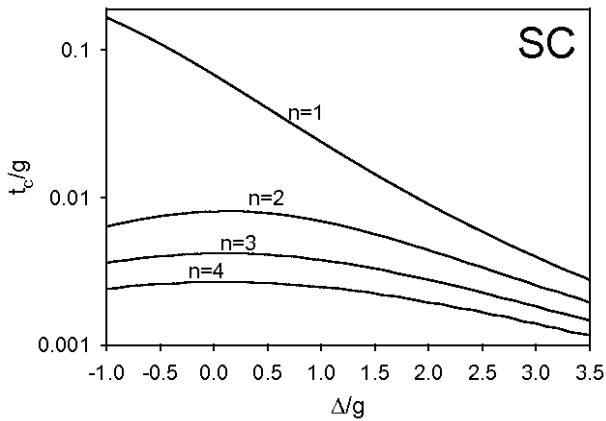}\label{fig3a}}
  \subfigure[]{
     \includegraphics[width=0.45\textwidth]{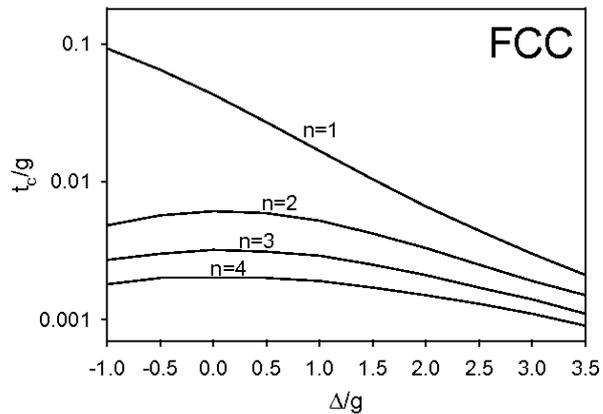}\label{fig3b}}
  \caption{Relationship between critical hopping and detuning on the (a) SC and (b) FCC lattices for varying excitation number.} \label{fig3}
\end{figure}

\begin{figure}
\includegraphics[width=0.45\textwidth]{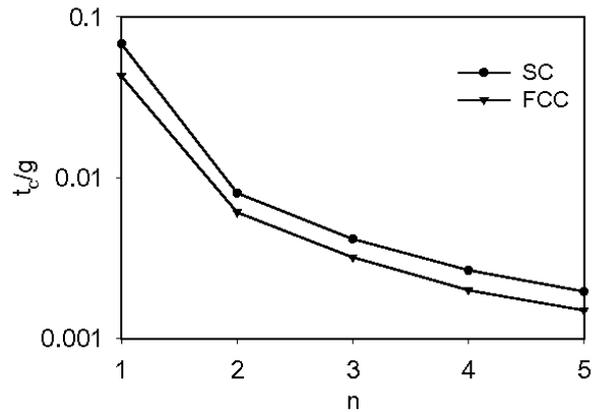}
\caption{Critical hopping for the SC and FCC lattices in terms of $n$ for null detuning.}
\label{fig4}
\end{figure}

\begin{figure}
\includegraphics[width=0.45\textwidth]{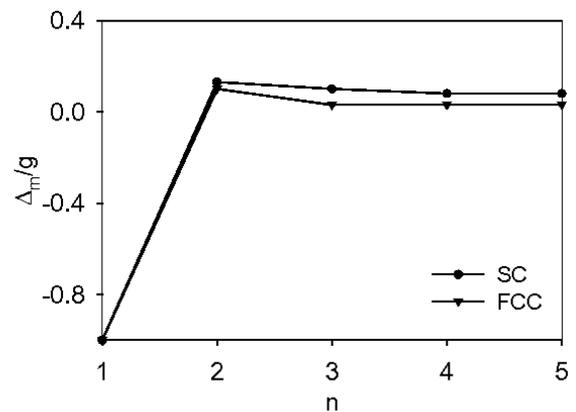}
\caption{Detuning versus $n$ for maximal critical hopping on the SC and FCC lattices.}
\label{fig5}
\end{figure}

The first Mott lobe ($n=1$) on the SC and FCC lattice is show in figure \ref{fig2} for typical detuning values.
We see that both lattices have a similar MI-SF phase transition frame. However, the FCC always has a smaller MI phase region.
For both lattices, the MI phase region decreases for increasing detuning.
Figure \ref{fig3} shows the critical hopping $t_c$ as a function of the detuning. While at the first lobe, $t_c$ decreases when $\Delta$ increases, in the other lobes ($n > 1$) 
we observe that the critical hopping reaches a maximum at $\Delta = \Delta_m$. 
For this particular detuning value, when $n$ increases the critical hopping decreases, as predicted in \cite{AMPK}. 
This behavior is present in both  structures.
Figure \ref{fig4} shows the critical hopping as a function of $n$ for $\Delta=0$. 
The properties expressed by both lattices are again qualitatively equivalent where there
is only one gap between the two curves. 
Figure \ref{fig5} presents the detuning values corresponding to the maximal critical hopping as a function of
$n$. We can observe that, except for the $n=1$ case, where the $t_c$ maximum is associated with the detuning minimum,
The critical hopping correspondent detuning is approximately null. 


\begin{figure}
\includegraphics[width=0.45\textwidth]{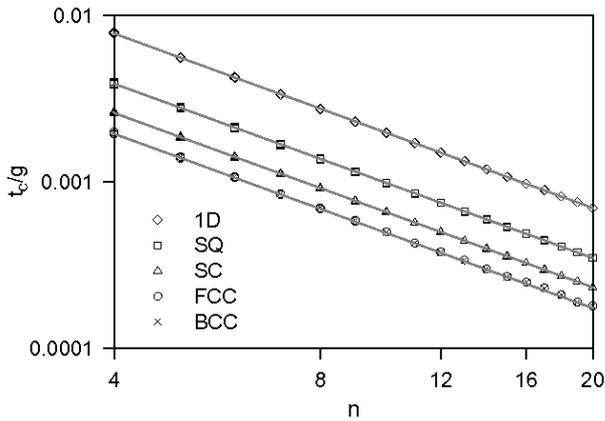}
\caption{Critical hopping versus $n$ for various lattices at the large $n$ regime. The symbols are related to 
numerically obtained results for
$\Delta / g = -0.5, 0,$ and $0.5$, where each detuning value produces the same result with errors smaller than the symbol size.
The solid line represents the asymptotic analytical result given by means of equation (\ref{tc.largen}).}
\label{fig6}
\end{figure}

By performing an asymptotic approach to large excitation number, $n \gg 1$, we can find the dominant term of the critical hopping which is given by
\begin{equation}
 t_c = \frac{g}{16 \tilde{d} n^{3/2}} + {\cal O} (n^{-3}), \label{tc.largen}
\end{equation}
where we obtain $\tilde{d}=4$ for the FCC and BCC lattices where it corresponds to the hypercubic lattices dimentions, i. e., 1 for linear, 2 for
square, and 3 for the SC lattice. It is important to emphasize that at the large-$n$ regime, the detuning 
and the topology class (bipartite or non-bipartite) influence on the critical hopping $t_c$ is suppressed.  
Figure \ref{fig6} confirms the prediction of the equation (\ref{tc.largen}). It shows that the
exact results are in excellent agreement with the asymptotic ones for $n>4$.

\section{Conclusions}
We have studied the properties of the MI-SF phase transition for the Jaynes-Cummings-Hubbard model over several Bravais lattices
by means of the fermionic approximation. We find that the transition parameters of hypercubic lattices are scalable, except for the FCC lattice,
since it is non-bipartite. 
The Mott lobes for the SC and FCC lattices show a similar detuning dependence
having only a quantitative difference which is suppressed as the detuning increases. 
An analogous feature is observed in $t_c$ vs. $n$ for null
delta and in $\Delta_m$ vs. $n$ where treir quantitative difference tends to be smaller as $n$
increases. 
Furthermore, we observed that not only the number of neighbors influences the
MI-SF phase transition but also does the lattice topology.
 
The FCC lattice shows a behavior quantitatively different and non-scalable from bipartite lattices. 
On the other hand, asymptotic
results for large excitation number indicate an universality on $t_c$ because it obeys a power law in $n$ which does not depend on
$\Delta$ and topology associated parameter can be rescaled for different classes through a multiplication of $t_c$ by an effective parameter
that corresponds to the dimension, for hypercubic, and to 4, for BCC and FCC lattices.

\section*{ACKNOWLEDGMENTS}
This work was partially supported by CNPq, CAPES and FAPITEC/SE (Brazilian Research Funding Agencies).

\end{document}